\documentstyle[twoside]{ms}   
\mysection{section}{\centering\normalsize\bf}{\thesection.~}
\mysection{subsection}{\centering\normalsize\it}{\thesubsection)~}
\mysection{subsubsection}{\centering\normalsize\rm}{\thesubsubsection)~}
\renewcommand{\thesection}{\arabic{section}}
\renewcommand{\thesubsection}{\thesection.\arabic{subsection}}
\renewcommand{\thesubsubsection}{\thesubsection.\arabic{subsubsection}}

\sfcode`A=1000 \sfcode`B=1000 \sfcode`C=1000 \sfcode`D=1000
\sfcode`E=1000 \sfcode`F=1000 \sfcode`G=1000 \sfcode`H=1000
\sfcode`I=1000 \sfcode`J=1000 \sfcode`K=1000 \sfcode`L=1000
\sfcode`M=1000 \sfcode`N=1000 \sfcode`O=1000 \sfcode`P=1000
\sfcode`Q=1000 \sfcode`R=1000 \sfcode`S=1000 \sfcode`T=1000
\sfcode`U=1000 \sfcode`V=1000 \sfcode`W=1000 \sfcode`X=1000
\sfcode`Y=1000 \sfcode`Z=1000

\newcommand{\js}{\rm}                             
\newcommand{\duprule}{\rule[0.5ex]{3.0em}{0.4pt}} 
\newcommand{\vol}[1]{{\rm{ #1}}}
\newcommand{\aj}{{\js AJ{\rm,}}}

\newcommand{\apj}{{\js ApJ{\rm,}}}

\newcommand{\apjs}{{\js ApJS{\rm,}}}

\newcommand{\aap}{{\js A\&A{\rm,}}}
\newcommand{\aaps}{{\js A\&AS{\rm,}}}

\newcommand{\pasp}{{\js PASP{\rm,}}}

\newcommand{\etal}{{\js et~al.\/}}
\newcommand{\eg}{{\js e.g.\/}}
\newcommand{\ie}{{\js i.e.\/}}
\newcommand{\micron}{\ifmmode\mu{\rm m}\else$\mu{\rm m}$\fi}

\newcommand{\ga}{\mathrel{\raisebox{-0.8ex}
        {\(\stackrel { \textstyle >} {\sim }\)} }}

\newcommand{\0}{\phantom{0}}  
\newcommand{\arcsec}{\(\stackrel{\:''}{\textstyle.}\)}

\newcommand{\arcdeg}{\(\stackrel{\:\circ}{\textstyle.\rule{0pt}{0.65ex}}\)}
\renewcommand{\deg}{\(^\circ\)}
\newcommand{\kms}{{\medmuskip=4.0mu plus 2.0mu minus 2.0mu
    \ifmmode\>{\rm km~s^{-1}}\else$\>{\rm km~s^{-1}}$\fi}}
\newcommand{\hi}{{\sc H~i}}
\newcommand{\hii}{{\sc H~ii}}
\newcommand{\tf}{Tully-Fisher relation}

\renewcommand{\thetable}{\arabic{table}}
\newenvironment{texttable}[1]   
       {\refstepcounter{table}
        \renewcommand{\thefootnote}{\alph{footnote}}
        \begin{center}TABLE~\thetable\\
        \medskip #1 \\ \smallskip
        \small
        }{\end{center}}
\newcounter{colcounter}[table]
\newcommand{\col}{\addtocounter{colcounter}{1}(\arabic{colcounter})}
\begin{document}

\hyphenation{in-cli-na-tion in-cli-na-tions gal-axy gal-ax-ies
sep-a-ra-ted adop-ted ap-prox-i-mate-ly cen-ter in-fra-red}
\title{The Infrared Tully-Fisher Relation\\ in the Ursa Major Cluster}
\author{R.~F.~Peletier \\
Kapteyn Laboratorium,\\ Postbus 800, 9700~AV Groningen, The Netherlands \\
and \\
European Southern Observatory, \\
K. Schwarzschildstr. 2, D-8046 Garching bei M\"unchen, Germany  \\
and
S.~P.~Willner\\
Harvard-Smithsonian Center for Astrophysics,\\
60~Garden Street, Cambridge, MA~02138 USA
}
\date{Accepted for publication in\\ {\it The Astrophysical
Journal\/}, 1993 December 1.}
\maketitle

\begin{abstract}

We present new  magnitudes derived from 1.65~\micron\ images for 23
galaxies in the Ursa Major cluster.  Magnitudes now exist for all but
one spiral meeting our criteria for cluster membership and having
\hi\ velocity width greater than 187\kms\ and inclination greater
than 45\deg.  These spirals fit a Tully-Fisher relation with
dispersion in intrinsic magnitudes (after known observational
uncertainties and the effect of cluster depth are removed) of 0.36
and a slope of 10.2 $\pm$ 0.6.  The magnitude dispersion is smaller
than found in the Virgo cluster but still significantly larger than
claimed by some authors.  We find a hint that the \tf\ may turn over
at the bright end.  Adding the central surface brightness of the disk
as a third parameter flattens the slope of the \tf\ and may give a
distance estimate with slightly less dispersion, but the significance
of the decrease must be tested on an independent sample.

\end{abstract}

\twocolumn
\section{Introduction}  

The \tf\ (Tully \& Fisher 1977) is one of the most useful ways to
measure distances to spiral galaxies (\eg, Jacoby \etal\ 1992).
However, the amount of scatter in the \tf\ is still a key issue,
one of importance not only for estimating the distance uncertainties
but also because the amount of scatter is crucial in estimating the
bias in the distances themselves (Teerikorpi 1984, 1987, Bottinelli
\etal\ 1987).  Several authors have found remarkably small
dispersions (\eg, Freedman 1990, review  by Jacoby \etal\ 1992), but Fouqu\'{e}
\etal\ (1990) and Peletier and Willner (1991, hereinafter Paper~1)
have found intrinsic dispersions among Virgo cluster spirals near
0.5~magnitudes in the blue and 0.4~magnitudes in the infrared.

The natural question is whether the large dispersions are a property
of just the Virgo cluster or are inherent in the \tf\ itself.  An
ideal test case is the Ursa Major cluster.  It lies at nearly the
same redshift as Virgo, has plenty of spiral galaxies, and earlier
studies (Pierce \& Tully 1988, hereinafter PT) have indicated a
smaller Tully-Fisher dispersion than in Virgo.

This paper presents magnitudes derived from infrared images for a
nearly complete sample of Ursa Major spirals.  The primary aim is to
investigate the dispersion, but we also examine how best to derive
magnitudes and inclinations for Tully-Fisher purposes.  Sample
selection is given considerable attention.


\section{Sample Selection}

Selection of the sample to be studied is crucial both to avoid biases
in the magnitudes, which could lead to bias in derived distances
(Teerikorpi 1984, 1987; Kraan-Korteweg, Cameron, \& Tammann 1988;
Bottinelli \etal\ 1988), and to calculate the correct dispersion in
the derived magnitudes (Paper~1).  An ideal sample would be selected
without any reference whatever to galaxy magnitudes.  Although the
ideal is impossible, for such nearby clusters as Ursa Major
a close approximation to the ideal can be achieved.

There are two methods commonly used to define cluster membership.
The first is based on a nearest-neighbor analysis (\eg, Huchra and
Geller 1982), while the second is simply to establish position and
velocity limits.  In the nearest-neighbor or ``tree'' analysis, a
list of galaxies with positions, velocities, and magnitudes is
somehow sorted according to the presumed ``closeness'' of the various
galaxies.  A density threshold is then established, and any section
of the list having density above the threshold is taken to be a
group. This method assumes no {\em a priori} knowledge of cluster
locations, but practical implementations (\eg, Tully 1987) often have
an explicit dependence on galaxy magnitudes.

The second method is necessarily somewhat empirical.  The position
limits are usually expressed as a cluster center and radius, and
minimum and maximum velocity limits are established.  The advantage
of this method is that there is no explicit dependence on galaxy
magnitudes, though a dependence could arise if the original list is
seriously incomplete at fainter magnitudes.  Incompleteness is not a
problem for this study, however, because catalogs contain galaxies
fainter than 15th magnitude (though they are not complete at this
limit, of course) while most of the spirals in the Ursa Major cluster
are brighter than 12th in~$B$.

For this study, we
examined data compiled for the CfA Redshift Survey (Huchra \etal\
1983 and unpub\-lish\-ed).  An initial examination in Right
As\-cen\-sion-Dec\-li\-na\-tion-Red\-shift space
found distinct groupings at $V_h
< 400 \kms$, $625 < V_h < 825 \kms$, $825 < V_h < 1025 \kms$, and $1100
< V_h < 1300 \kms$.  The last group seems to be separated from the
others in position, while the first three are separated mainly in
velocity.    The position centroid of those galaxies in the
range $600 < V_h < 1050 \kms$ is approximately 11$^H$~54$^M$,
48$^\circ$~53$'$, which we have adopted as the cluster
center.\footnote{
  Compare with 11$^H$~54$^M$, $+49^\circ$~30$'$ found by Biviano
  \etal\ (1990).  PT used the same center coordinates as Biviano
  \etal\ along with a cluster radius of 7\arcdeg5 and effective
  velocity limits $628 < V_h < 1138 \kms$.}
The surface density of galaxies falls off beyond 7$^\circ$ from this position,
so we have adopted this as the cluster radius even though there are
undoubtedly cluster galaxies beyond this radius.  Finally, we have
slightly extended the velocity limits to $550 < V_h < 1150 \kms$ in
order to make our definition correspond more closely to previous work.
All galaxies meeting these requirements are listed in either
Table~\ref{tab:sam} or~\ref{tab:exc}.\footnote{
  All galaxies in Tables~\ref{tab:sam} and~\ref{tab:exc} except two
  are cataloged galaxies for which positions and other data can
  be found in the NASA Extragalactic Data\-base (NED).
  115400+4836 (KDG~310A) is a companion
  south preceding NGC~3985 (KDG~310B).  We do not know of published
  data on 115640+5059.  These two galaxies are among the ones excluded
  from  our Tully-Fisher sample and were not observed.}
There is no explicit magnitude dependence in selecting this sample,
or indeed any requirement that magnitudes be known, but galaxies must
have been cataloged and had redshifts measured.  This will introduce
some incompleteness, but this probably becomes serious only below
magnitude~14.

 From the initial list, we have excluded 15 galaxies that are not
spirals (keep\-ing only galaxies with $T>0$ according to the RC3---de
Vaucouleurs \etal\ 1991).
Only reasonably edge-on galaxies can give useful
velocity widths, so we have excluded 13 galaxies with axis ratio less
than 1.40 corresponding to inclination less than 45$^\circ$
(Paper~1).  Finally, we have excluded 17 galaxies that lack \hi\
observations or have inclination-corrected velocity widths $\Delta
V^c_{20} < 187.5$\kms, an arbitrary limit reflecting the
completeness of our photometric observations.  Many of the excluded
galaxies are faint, but it is possible that better \hi\ observations
would allow some of them to be used for Tully-Fisher purposes.
Table~\ref{tab:exc} lists all the excluded galaxies.

The remaining cluster sample contains 29 galaxies listed in
Table~\ref{tab:sam}.  In practice, an almost identical sample would
have been obtained from nearest-neighbor analysis sim\-ply by adopting
group ``12-1'' from Tully (1987).\footnote{
  The only differences affecting Table~\ref{tab:sam} are that Tully
  (1988, hereinafter NBG) assigns
  NGC~3985 and 4096 to group 14-4.  The former galaxy
  has an incorrect heliocentric velocity in the NBG, and it seems it
  would have been considered a
  member of 12-1 if the correct velocity had been used.  The latter
  is on the outskirts of both groups and might belong to either.}
This group has 57 members, and other authors of nearest-neighbor
analyses (Geller and Huchra 1983, Huchra and Geller 1982, Turner and
Gott 1976) consider it a sub-group of their somewhat larger groups.
The difference seems to arise because the latter authors are
interested in larger structures and accordingly have set their
density thresholds lower.  For this study, we need to be sure all the
galaxies are at the same distance, so the most restrictive definition
is appropriate.

Two of the galaxies in Table~\ref{tab:sam} lack infrared magnitudes.
UGC~6917 has a bright star superposed on the galaxy, rendering the
current observations useless, and we failed to observe NGC~4218.  A
total of 27 galaxies are thus available for analysis.


\section{Observations}  

All of the observations were made with the 1.2-m telescope at
Mt.~Hopkins and the Smithsonian Observatory Near Infrared Camera
(SONIC). This camera is the same one used with the 0.6-m telescope
for Paper~1. 20 galaxies were observed in 1991 and 4 in 1992 through
a standard (Barr Associates) 1.65~\micron\ (``$H$'') bandpass filter.
Observations in 1991 January through March used a two-lens reimaging
optical system giving a scale of 1\arcsec641 per pixel, while those
in 1992 March used a three-lens optical
system giving 1\arcsec757 per pixel.    Sky frames were
observed along with each galaxy, separated from object
frames by $>$3~arcmin depending on the galaxy size.

Data calibration began by removing dark current and star images from
sky frames and averaging sky frames to create a flat field.  Each
object frame, including those for standard stars, was divided by a
flat field frame.  The observations of standard stars showed a
variation of up to 30\% peak-to-peak in 1991 and 20\% in 1992
depending on where the standard star fell on the object frame.  The
variation was repeatable and mostly in the East-West direction.  We
attribute it to the incomplete long-wavelength blocking of the
1.65~\micron\ bandpass filter; the sky frames thus contain a
component at $\sim$5~\micron, while the much bluer stars contain no
such component.  Since the quantum efficiency of the detector is a
function of wavelength as well as position, the incomplete blocking
leads to the sky flats being inaccurate when applied to stars.  The
change from 1991 to 1992 is attributed to addition of a glass
blocker, but obviously a thicker blocker is needed (and has since been
added).

Fortunately, it is easy to calibrate the responsivity variation since
nearly all standard stars were observed at many positions on the
detector array.  A linear function of pixel x-y coordinates was
derived for each night, though in practice there was no significant
variation from night to night.  A normalized sky frame was subtracted
from each object frame, then the result was multiplied by the
correction frame.  This procedure should work well because the
standard stars are nearly the same color as the galaxies.  Finally,
the various object frames were averaged and mosaiced together.

Typical final galaxy frames consist of images taken at 2 to 5 positions,
mosaiced with usually the galaxy nucleus as a reference point or
sometimes a bright star or an \hii\ region. Figure~\ref{fig:typ} shows
gray scale images of a bright, intermediate, and faint galaxy in the
sample. The dark current fluctuations mentioned in Paper~1 had been
cured, so the present observations reach typical 1$\sigma$ noise
levels of 20.7~$H$~mag~arcsec$^{-2}$
for bright galaxies and 21.2~$H$~mag~arcsec$^{-2}$ for the
faintest galaxies.  These values imply that our photometry is limited
by the accuracy to which the sky background can be determined on the
frames.  This in turn is limited by the small field size and by the
extent to which the images could be corrected for the unblocked
long-wavelength light.  By looking at the dispersion in the sky
values measured at various corners on the frames, these uncertainties
have been estimated and are given in Table~\ref{tab:mag} for the
various types of magnitudes.

On 9 out of 11 nights the weather conditions were photometric.  On
these nights the photometric zero points were calibrated using
standard stars from Elias \etal\ (1982). Typically on each night 5--6
red and blue standard stars were observed at various positions on the
frame. Internal consistency was better than 0.02 mag. No color term
was applied to the calibration.  On the other two nights the zero points
were calibrated using aperture photometry from Aaronson \etal\ (1982;
hereinafter A82).

Given the fact that the galaxy frames were well behaved and of a much
better quality than those for Virgo in Paper~1, accurate surface
brightness profiles could be determined.  Radial profiles of surface
brightness, ellipticity, and position angle were determined using
GALPHOT, the two-dimensional ellipse-fitting package written by
M.~Franx  (J\o rgensen, Franx, and Kjaergaard 1992).
Not only did we determine from them the axis ratio and the
position angle in the outer parts, to get the infrared inclinations,
but also a bulge-disk decomposition was performed. Analysis of the
photometric profiles will be given in a later paper, but
we will deal with whether bulge-disk decomposition can improve the
\tf\ in Section~4.


\section{Ingredients for the Tully-Fisher relation}

%

\subsection{Magnitudes in circular beams}

Since infrared arrays have only recently become available, people up
to now have used magnitudes in circular beams for the infrared
Tully-Fisher relation. Aaronson and co-workers (Aaronson, Huchra, \&
Mould 1979, hereinafter AHM; Aaronson, Mould, \& Huchra 1980,
hereinafter AMH; Aaronson \etal\ 1986, Bothun \etal\ 1985)
have used the magnitude inside circular beams with diameters of $0.316
D_1$, in which $D_1$ is the isophotal diameter at $B =
25$~mag~arcsec$^{-2}$
corrected for Galactic extinction and inclination
(see AMH).  To test the reliability of our photometry we have
determined the magnitudes inside $0.316 D_1$ for the galaxies that
overlap between A82 and this paper. The comparison is shown in
Figure~\ref{fig:a82}.  The mean difference, 0.02 mag, and the
dispersion, 0.03 mag, are well within the uncertainties of our data,
even for the faintest galaxies.  It shows that, if anything, our
photometric uncertainties have been overestimated.

Table~\ref{tab:mag} (column~3) lists circular magnitudes
($H^c_{-0.5}$) for sample galaxies in diameters of $0.316 D_0$.  Here
$D_0$ is the  blue isophotal diameter again at
25~$B$~mag~arcsec$^{-2}$ but now corrected for
Galactic extinction and inclination following the recipe of the RC2
(de Vaucouleurs \etal\ 1976),
analogous to Paper~1. $D_0$ is similar to $D_1$ and in  fact
is just $1.05 D_1$  for our galaxies. The inclination correction  is
meant to make sure that the same fraction of the infrared light
is measured for each galaxy  and corrects for the fact that the
ratio of effective to isophotal radii decreases with inclination for
transparent galaxies.  The inclination correction was not
applied to blue magnitudes in the RC3, since its authors were convinced
that spiral galaxy disks have surface brightness independent of inclination
in~$B$.  For $H$ magnitudes, however, the correction must
be applied, since in this band spiral galaxies are more or
less transparent (Peletier and Willner 1992).\footnote{
    We make no representation
  that this is the best possible method of determining a diameter, but
  the prescription is well defined, the method is commonly used, and no
  other method has been shown to be better (\S\ref{sect:which}).}
For those galaxies in
Ursa Major that are included in A82 for which we don't have infrared
images the circular magnitude inside $0.316 D_0$ was calculated from
the value inside $0.316 D_1$ using the average difference for the
galaxies in common ($-$0.03~mag).  These galaxies can be identified
by the absence of additional magnitudes in Table~\ref{tab:mag}.

\subsection{Magnitudes in elliptical beams}

An advantage of imaging is  flexibility in choosing
the kind of magnitudes to use. To investigate whether the scatter in
the Tully-Fisher law might decrease using different magnitudes we
have determined magnitudes in elliptical beams within the optically
determined $0.316 D_0$ as well as elliptical magnitudes in diameters
determined from infrared isophotes. For the latter, no information
from the optical is needed. Defining $D_{\mu,i}$ as the major axis
diameter of the isophote at surface brightness~$\mu$, corrected for
inclination using the recipe of the RC2 with infrared axis ratios, we
have calculated for each galaxy the magnitudes inside $D_{19,i}$ and
$0.7 D_{20,i}$, to be called $H_{19}$ and $H_{20}^{0.7}$. We used $0.7
D_{20,i}$ instead of $D_{20,i}$ because of the limited field of
some images.  These magnitudes have also been tabulated in
Table~\ref{tab:mag}.

\subsection{Total magnitudes}

Even though we never cover the entire galaxy, it is possible, with
some assumptions, to derive a good approximation for the total
magnitudes.  The reason is that the spiral galaxies of Ursa Major can
be fit rather well by a central bulge and an exponential disk. Total
magnitudes can be found simply by extrapolating the disk outward.
Schommer \etal\ (1993) have examined this method for $I$-band images
and emphasized its difficulties: because galaxy surface brightness
profiles show bends and wiggles, the uncertainty of any extrapolation
is increased and difficult to estimate.  There is little doubt that
deeper images would be preferable, but the uncertainties in the total
magnitudes derived this way are still less important than the
uncertainties in velocity widths and inclinations.

We have performed the bulge-disk decompositions using the method
described by Kent (1986). This method uses the surface brightness
profiles on the major and minor axes and assumes that bulge and disk
both have a constant axis ratio with the bulge being rounder than the
disk.  After the decomposition, we fit an exponential to the surface
brightness profile of the disk,  excluding the inner areas in
which the bulge dominates. The luminosity of the disk is calculated
analytically,\footnote{
  For an exponential disk with major axis profile $H(r)
  = H(0) \exp(-r/h)$ and major to minor axis ratio $a/b$, the
  integrated luminosity is  $2\pi (b/a) {h^{2}} H(0)$.}
and the bulge luminosity is
determined on the frame of each galaxy after having subtracted the
model disk. Since for almost all galaxies the bulge is much smaller
than the disk, the decomposition is unambiguous, and the
uncertainties in
the total magnitudes should be comparable to those in the circular
magnitudes.  Table~\ref{tab:buldis} gives the
ellipticities of bulge and disk, total magnitudes, bulge to disk
ratios, scale lengths, central disk surface brightness, and total
magnitudes of the disk.

\subsection{Which magnitudes are best?}
\label{sect:which}

Each type of magnitude discussed
above has been fit to a linear  \tf, taking into account
uncertainties in both magnitudes and velocity widths.  Since
the latter are so much larger than the former,
the procedure is almost equivalent to using magnitude as the
independent variable, \ie, to the ``inverse \tf'' discussed by Fouqu\'e
\etal\ 1990. The results are shown in Figure~\ref{fig:ell}
and Table~\ref{tab:lsq}, where
column~6 gives the reduced chi-square of the fit and column~7 shows
the additional magnitude uncertainty that must be added (in
quadrature) to make $\chi^2_{red} = 1$.

The smallest scatter is found for circular magnitudes, in agreement
with Paper~1.  The scatter increases only slightly for elliptical
magnitudes but is much greater with infrared isophotal magnitudes.
These have the advantage that they can be obtained without needing
optical images, but they do not give very satisfactory results. The
problem is that both bulge to disk ratios and disk surface
brightnesses decrease as a function of velocity width, which means
that some galaxies barely reach the isophote of 19~mag~arcsec$^{-2}$.
The \tf\ with these magnitudes thus displays a large amount of
curvature, as shown in Figure~\ref{fig:ell}, with especially the
faintest galaxies seeming much too faint for their velocity widths.
Optical isophotal magnitudes or infrared magnitudes within optically
determined diameters work much better because of the much smaller
scatter in the central surface brightness of disks in $B$ as opposed
to $H$ (Freeman 1970, Peletier \& Willner 1992).

With total disk magnitudes or total magnitudes the scatter is
slightly higher than when using circular magnitudes. Although in most
cases bulge to disk ratios are small and exclusion of a bulge does
not affect significantly the residual from the \tf,
NGC~3718 is different.  This galaxy deviates
considerably if its bulge is excluded. At least for this galaxy,
the \tf\ does not purely involve the disk but rather the whole galaxy.

In what follows, we adopt the magnitudes in circular beams.

\subsection{Velocity widths}

All velocity widths used here are H~{\sc I} velocity widths from
Bottinelli \etal\ (1990), a large compilation of the
literature. Since uncertainties in the velocity widths are
unimportant compared to uncertainties in inclination, and since
the errors for Ursa Major are on the average 8 km~s$^{-1}$, compared
to 13.5 km~s$^{-1}$ in our Virgo sample (Paper~1), we have made no
further attempt to select the best individual observations.

The only question in velocity widths is whether it is important to
correct them for non-rotational motions (Tully \& Fouqu\'e 1985).
The correction affects only the faintest galaxies.
Table~\ref{tab:lsq} shows that replacing $\Delta V^c_{20}$ by
corrected widths $W_R$ (Tully \& Fouqu\'e 1985) does not improve the
quality of the fit either for circular or for total magnitudes.

In what follows, we have used  widths $\Delta V^c_{20}$ corrected
only for inclination and not for non-rotational motion.

\subsection{Inclinations}

Paper~1 found that the best way to determine galaxy
inclinations is to use optical axis ratios in the outer parts.
Inclinations determined in the infrared often suffer from central
bars or strong spiral arms, because of the small field. However,
infrared inclination might turn out to
be better than optical ones if the surface
photometry is deep and the field large enough.

Even though the surface photometry for this paper goes approximately
2~mag deeper than in Paper~1, the agreement between infrared and
optical inclinations is no better.  Figure~\ref{fig:incl} shows the
difference between the inclinations.  Here an intrinsic axis ratio of
0.15 was assumed in the blue as well as the infrared (cf.\ Bottinelli
\etal 1983) because several galaxies were inconsistent with the
conventional value of 0.20 (\eg, AHM, Helou, Hoffman, \& Salpeter
1984).  The infrared inclinations confirm that
some galaxies have intrinsic thickness smaller than $b/{a}=0.20$
in this band as well.
The comparison between infrared and blue inclinations here is
qualitatively different from Paper~1. Although the agreement for 16
out of 23 galaxies is better than 4 degrees, the others show very
large differences in both directions. In most cases the difference is
caused by a faint envelope visible on optical plates, but for
NGC~4102 and NGC~4217 the differences are hard to explain this way.

Using infrared inclinations in the \tf\ increases rather than
decreases the scatter for all types of magnitude.  Most of this is
due to NGC~4389, which has a large residual to the \tf. Since its
infrared inclination is large, this will increase the residual and
decrease at the same time the apparent uncertainty in $\Delta V^c$,
making the scatter much worse.  However, even without NGC~4389 the
scatter is larger when using infrared inclinations. For the subsample
in common with PT, the scatter decreases if CCD inclinations are
used.  It might therefore be worthwhile to obtain CCD inclinations
for the whole sample.

PT obtained inclinations for many Ursa Major galaxies using CCD
photometry. The average difference between the inclination derived by
PT and from the RC3 is 1.8$^\circ$ with a scatter (rms) of
4.6$^{\circ}$. The situation here is similar to that in Virgo, for
which we claimed (Paper~1) that the uncertainty in the inclination of
a typical galaxy is 5 degrees.  Intrinsic uncertainties, like an axis
ratio that varies with radius, spiral arms in the outer regions, and
slightly triaxial shapes (Franx \& De Zeeuw 1992) make it very
difficult to see how this uncertainty can be reduced.  Schommer
\etal\ (1993) derived inclinations both photometrically ($I$-band)
and from kinematic fits to the velocity field of the gas and thereby
produced Tully-Fisher relations for two clusters with a {\em total}
scatter of $<$0.30~mag.  Kinematic inclinations or a
combination of photometric and kinematic inclinations thus seem quite
promising.  However, the two methods give large ($>$10\deg)
differences in inclination for some galaxies, and caution seems
advisable until these differences can be understood.

\subsection{Type dependence}

No dependence of the \tf\ on galaxy type was found,
in agreement with Paper~1.

\subsection{More parameters to reduce the scatter}

With the advantage of having images, we have investigated whether the
residuals from the \tf\ correlate with other galaxy parameters.
Although for most parameters the correlation is weak,  the
central disk surface brightness may have a useful effect.
Table~\ref{tab:bivar} gives the results of fitting a plane to the
data in the space of magnitudes, velocity widths, and central surface
brightness.\footnote{
  The uncertainties in central surface bright\-ness are not
  important, since this is a second order correction, so only
  uncertainties in magnitudes and velocity widths have been considered.
  The fit chosen was the one that minimizes the additional magnitude
  uncertainty needed to give a reduced chi-square of one, not the one
  that minimizes chi-square itself.  The fit is rather insensitive to
  the exact coefficient of the $H(0)$ term with values between 0.2 and
  0.5 giving about the same result.}
Although the scatter measured by chi-square does not decrease (in
fact increases slightly), the slope of the relation flattens, and a
given uncertainty in velocity width translates to a smaller
uncertainty in magnitude.  Figure~\ref{fig:biv} displays the results
graphically.  The resulting scatter is, however, still too large to
be explained entirely by the depth of the cluster.  Since a variety of
other galaxy parameters might have been (and were) tested for their
ability to reduce the dispersion, the usefulness of this one must be
verified on an independent sample before being accepted.


\section{Changing the Sample}

Having found which input data give a \tf\ with the least scatter, we
here investigate whether changing the sample can affect the
conclusions. Table~\ref{tab:pt} and Figure~\ref{fig:baseline} show the
results for the complete sample given in Table~\ref{tab:sam}.  The
principal result is that even after consideration of the known
uncertainties in the observations, there remains uncertainty in the
magnitude of an individual galaxy of 0.36~mag once the effect of the
expected cluster depth (0.17~mag) is removed.  This is in agreement
with the value found for Virgo in Paper~1.

\subsection{Previous work}

PT obtained CCD images in $B$, $R$, and $I$ for 26 Ursa Major
galaxies and obtained total magnitudes in these bands.  For 18 of
these, $H^c_{-0.5}$ magnitudes had been obtained by A82.  PT found a
scatter around the \tf\ of $\sim$0.30 mag in $R$, $I$, and $H$.
Corrected for uncertainties in the observations, they needed an
intrinsic scatter of only 0.22 mag, which is statistically consistent with
the depth of the cluster ($\sim$0.17~mag if the cluster is a sphere).
However, the sample by PT is not complete, although for $B_T<
13.3$~mag it contains most cluster members.\footnote{
  The PT sample contains five galaxies that we rejected from our
  initial sample (Table~2) and one more (UGC~6816) that is 8$^\circ$
  from our adopted cluster center.  However, of these rejected
  galaxies, only NGC~3782 has $H$~band photometry.  Thus the main
  difference between our samples is that we have added eight galaxies
  (plus NGC~4218 which we didn't observe)
  not included by PT.  As noted, we have also measured two galaxies that were
  included in their sample but had no $H$~photometry.}

As in Paper~1, we have re-analyzed their sample with new velocity
widths and magnitudes.  Table~\ref{tab:pt} compares Tully-Fisher fits
for our sample and for the PT sample. The first line of the PT
results uses only their data except that the older magnitudes have
been replaced by ours.  Successive lines show the effect of including
the two galaxies not previously measured, using RC3 velocity widths,
removing the velocity width correction for non-rotational motions,
and using inclinations derived from the RC3 instead of the PT
inclinations.  None of these changes makes any difference in the
scatter as measured by $\chi^2$, although the last one does
increase the inferred magnitude
uncertainty.\footnote{
  Table~\ref{tab:pt} also shows that omitting NGC\penalty500\ 3782, the only
  galaxy in the PT sample but not in ours, makes little difference.}
Figure~\ref{fig:ptcom} shows the \tf\ derived from both
sets of inclinations.  Almost all of the increased magnitude
uncertainty comes from UGC~6983, to
which PT assign an inclination of 55\deg\ while the RC3 axis ratios
imply $i\approx47^\circ$.  Nevertheless, for this sample, for all
choices of velocity widths, correction procedures, and inclinations,
the intrinsic scatter after correcting for observational
uncertainties is not much more than 0.2 mag, in agreement with that
found by PT. The additional scatter found in our complete sample must
therefore be contributed by the galaxies not included in the PT
sample.  This increased scatter is inconsistent with the expected
cluster depth and may be regarded as intrinsic scatter in the
magnitude of an individual galaxy.  Some possible causes of the
scatter were discussed in Paper~1.

\subsection{Outlier Galaxies}

As shown in Figure~\ref{fig:baseline}, the two galaxies with the largest
residuals are among the eight absent from the PT sample.  Six of the
eight galaxies have residuals larger than the 1$\sigma$ value, as
opposed to 3 one would expect from Gaussian statistics.  The four galaxies
causing most of the extra scatter are the two bright galaxies, NGC~4389, and
possibly UGC~6894. It is debatable that PT did not include NGC~3718,
since in the RC3 it has been classified as peculiar or a merger
remnant, but there is no obvious reason why the brightest cluster
member, NGC~3992, was not included.

NGC~3718 has a regular \hi\ profile and does not look disturbed in
the $H$~band. However, in the optical a prominent dustlane is
visible, together with faint structure in the outer regions.  The gas
distribution can be explained well by a warped disk with a tilt of
almost 90$^\circ$ (Schwarz 1985).  Given this extra information that
the dynamics of this galaxy are dissimilar to a rotating disk, one is
justified in removing it from the sample. However, if only the
minimum information required to use this galaxy in the \tf\ were
available, as is usually the case, one would have no reason to reject
it.  Changing the inferred inclination of the galaxy would not bring
it closer to the \tf. The optical axis ratio from the RC3 was taken
in the outer regions, where the galaxy is the most elongated, so that
the inclination correction is minimized.  An inclination derived from
photometry in the inner parts (\eg, the H-photometry) would only
increase the deviation.  If we reject NGC~3718 from the sample we see
that the scatter decreases only slightly (Table~\ref{tab:pt}).

Inclusion of NGC~3718 and 3992
makes it appear that the \tf\ levels
off at large velocity width, \ie, that above a velocity width of
$\sim$450\kms\ galaxies have a constant brightness rather than a
further increase.  The turnover\footnote{
  To avoid confusion, we use the word ``turnover'' for possible
  leveling off at the bright end of the \tf.  It should not be
  confused with ``curvature'' at the faint end.  The latter can
  always be eliminated by a suitable choice of correction for
  non-rotational velocities.}
is also seen in the data
presented by AHM (Figure~2), PT (Figure~3), and Giraud (1986,
Figure~5), although these authors did not comment on it.  No turnover
is seen, however, in observations of the Coma cluster (Raychaudhury,
private communication).

The deviation of NGC~4389 might be caused by an underestimated
velocity width.
Contrary to most of the other galaxies in his sample, the velocity
profile of NGC~4389  is not symmetric (Huchtmeier 1982). Its type,
Sa, is another indication that this galaxy might be deficient in \hi.
Table~\ref{tab:pt} shows the effect on the \tf\ of omitting NGC~4389.
The scatter goes down, but the remainder is still too large to be
caused  by  cluster depth.  Even
omission of both NGC~3718 and 4389 does not reduce the scatter to the
level found by PT.

We have also considered the possibility that for the faintest
galaxies, the systematic errors are much larger than assumed.  Since
even in $B$, surface brightness decreases as a function of velocity
width, the circular magnitudes might not be adequate any more;
systematic errors in the bulge-disk decomposition might equally well
affect the total magnitudes.  Table~\ref{tab:pt} shows results for a
sample of galaxies with $\Delta V^c$ $>$ 2.43. The residuals from the
\tf\ here are smaller than for the complete sample but still closer
to 0.4 than to  0.2 mag.  The scatter remaining is caused by
the turnover at the bright end of the relation.

Excluding NGC~4096, the one galaxy in our sample correctly assigned
to group 14-4 instead of 12-1 (NBG), makes no significant
difference in the results.

\subsection{Evidence of sub-clustering}

Although our definition of ``the Ursa Major Cluster'' has closely
followed previous work, one interpretation of our data is that there
is a background sub-cluster superposed on the main cluster.
Figure~\ref{fig:vel} shows the residuals from the \tf\ plotted as a
function of heliocentric radial velocity.  The four galaxies with large
residuals (in the sense that they are fainter than expected for their
velocity widths) all have radial velocities $\ga$1000\kms.  Galaxies
with high velocity (including rejected galaxies) are mostly located
north of the cluster center, though some are found to the south as
well.  The data of PT show no evidence for a background sub-cluster,
nor are we aware of any previous suggestion of one, but the
possibility must be mentioned.

If galaxies with radial velocity great\-er than 1025\kms\ (or less
than 600\kms) are excluded from the sample, the derived intrinsic
scatter drops to 0.26~mag (Table~\ref{tab:pt}.  This approaches the
amount of scatter from the expected cluster depth. If the upper
cutoff is reduced to 985\kms\ so as to exclude NGC~3718, the
intrinsic scatter drops to 0.12~mag, significantly less than
expected.


\section{Discussion}

\subsection{Intrinsic scatter}

In the previous section it was found that for a complete sample in
Ursa Major, selected in $\Delta V^c$, the scatter in the \tf\ is larger
than can be explained by known observational uncertainties or  the
depth of the cluster, assuming it is spherical. Since PT obtained a
dispersion that was consistent with no intrinsic scatter, and our
samples differ basically in the brightest and faintest
galaxies, it is possible that the scatter originates from
curvature in the relation rather than from intrinsic scatter at each
velocity width.  Even when removing the faint galaxies ($\Delta V
\le 2.43$), for which the observational uncertainties
might be larger than estimated,
the scatter remains 0.31~mag after correcting for observational
uncertainties.  A turnover at the bright end of the \tf\ in Ursa
Major is a possible explanation.  Curvature at the faint end of the
\tf\ was discussed earlier by A82, Aaronson \& Mould (1983),
Bottinelli \etal\ (1984), and others. Bottinelli \etal\ pointed out that
this could be an artifact of the line width definition, and
PT found no curvature when using $W_R$ instead of $\Delta V^c_{20}$.
However, Table~\ref{tab:lsq} shows that correcting for non-rotational
motion has no significant effect on the scatter.

Two recent studies of the $I$-band \tf\ (Mathewson \etal\ 1992,
Schommer \etal\ 1993) obtained values for the scatter that are
much lower than we find for Ursa Major.\footnote{
  Schommer \etal\ found a scatter of 0.29~mag for Hydra and only
  0.18~mag for Antlia.  Matthewson \etal\ found 0.25~mag for Fornax.
  These values appear to represent the {\em total} scatter, \ie, the
  scatter including that due to  observational uncertainties.}
However, neither  of these studies claim to be
complete in any sense, and Schommer \etal\ observed very few cluster
members.  If more detailed study shows such low scatter to be real,
the sub-structure in Ursa Major must be taken seriously.

In this paper we found an enormous amount of curvature when using
magnitudes defined using infrared isophotes. Since the disk surface
brightness decreases rapidly as a function of $\Delta V$
(Table~\ref{tab:buldis}), the ratio
between isophotal and total magnitude changes with it in a non-linear
way.  In $B$, where the diameter used for $H^c_{-0.5}$ is determined,
the central surface brightness is roughly constant for bright
galaxies but decreases with luminosity for faint galaxies (\eg,
Peletier \& Willner 1992). Since the latter effect is much weaker
than in $H$, the use of $H^c_{-0.5}$ might result in some subtle
curvature.  This cannot be the cause of the observed scatter,
however, because the scatter remains even if faint galaxies are excluded
or if velocity widths are corrected for non-rotational motions.

The turnover at large $\Delta V$ remains if we use total magnitudes
but can be removed by taking the central surface brightness of the
disk, $H(0)$, as a third parameter (Figure~\ref{fig:biv}).  Even with
this third parameter, the magnitude dispersion is still too large to
be explained by the observations alone. The dependence of magnitude
on $\Delta V$ and $H(0)$ may be non-linear, not only the $\Delta
V$-dependence (the curvature), but also the $H(0)$-dependence.
Bothun \& Mould (1987) suggested that by using the surface brightness
profiles (not just central surface brightness), they could decrease
the scatter in the \tf.  If non-linear relations are allowed, many
more galaxies must be observed to determine the additional free
parameters.  In any case, there seems no way of avoiding significant
intrinsic scatter in the conventional {\em linear} \tf.

\subsection{The slope of the Tully-Fisher relation}

Depending on the choice of magnitude and sample, most of the slopes
we find lie between 0.09 and 0.11 (for $\log \Delta V^c = -aH + b$),
corresponding to a conventional slope ($H = -a'\log\Delta V^c + b'$) between
9 and 11. This is consistent with the Virgo cluster (Paper~1),
although that sample was magnitude selected and therefore potentially
biased in slope. PT show that for $B$, $R$, and $I$ the slope using total
magnitudes is smaller than when using aperture magnitudes. We find
the same in $H$ ($9.3 \pm 1.0$  vs.\  $10.2 \pm 0.6$) for $H_T$ and
$H^c_{-0.5}$ for our largest possible samples.  However, the slope for total
magnitudes is not 8, as PT predict, but close to 10, a value that is
expected for $\Delta V^c \propto M^4$ (as expected if the central
mass surface density of galaxies is constant)
and a constant $M/L$ (AHM 79). Since the
total magnitudes in this paper have all been determined using
extrapolation, this conclusion will have to be tested with deeper
surface photometry.

\subsection{Difference between Ursa Major and Virgo}

Although we find significantly larger scatter in both clusters than
did PT, we confirm their result that the scatter in Ursa Major is
smaller than in Virgo. From this it follows that there must be some
substructure in Virgo or that the Virgo cluster is elongated towards
us, or the Tully-Fisher intrinsic scatter varies from cluster to
cluster.  Our zero points for Virgo and Ursa Major
are almost identical (for the ``baseline'' 2.547 in Virgo and 2.564 in
Ursa Major), so it appears that the center of Virgo is $8 \pm 3$\%
closer than Ursa Major, in agreement with PT.  However, this
conclusion should be regarded with caution, since the Virgo sample
was magnitude selected and therefore potentially biased in magnitude.

\subsection{Extinction in H}

Even in $H$, galaxies might still contain reasonable amounts of
extinction (Peletier \& Willner 1992).  A typical face-on extinction
of 0.07~magnitudes corresponds to 0.20~mag for a galaxy with
inclination 70$^\circ$, typical of this sample.  Bothun \& Mould (1987)
derived a typical reddening-correction to their $I$~data of 0.30~mag.
Using the galactic extinction law (Schultz \& Wiemer 1975) this
corresponds to 0.11~mag in $H$. Either amount would be enough
to cause measurable scatter in the Tully-Fisher relation. We have
found no correlation between inclination and total H-magnitude, but
our sample is small, and our total magnitudes are uncertain.  There
is also no correlation between inclination and residual other than
the expected greater scatter for lower inclinations.  Since the
reddening is very uncertain, a study of the infrared color profiles
of spiral galaxies is important to determine whether absorption
causes any of the scatter in the Tully-Fisher relation, even in~$H$.

Freeman's (1970) law, which states that the central surface brightness of
disks of spiral galaxies of types Sa--Sc is constant at $B=21.6$
mag~(arcsec)$^{-2}$ with a scatter of only 0.3~mag, implies that the
centers of some disks are very red in $B - H$. Some galaxies have
central disk colours of $B-H=6.0$~mag.  Since a typical elliptical
galaxy (thus presumably a typical spiral bulge) has a $B-H\approx
3.9$, there has to be at least 2.5~mag of extinction in $B$ or
possibly more, depending on the geometry.  Many galaxies of this
sample are so red in the center that they are optically thick in $B$.
The colors will be discussed further elsewhere.


\section{CONCLUSIONS}

 From a H-band imaging study of a complete sample of galaxies in the
Ursa Major cluster, selected on the basis of velocity width we
conclude:

\begin{itemize}

\item The scatter in the \tf\  is least when using circular
magnitudes within an optically determined diameter.
Infrared isophotal magnitudes
give very bad fits owing to the strong
relation between velocity width and surface brightness. Using
extrapolated total magnitudes the scatter is worse than using
circular magnitudes.  It is still possible, however, that total
magnitudes measured from deeper images may match or improve upon the
circular magnitudes.

\item The scatter in the \tf\  in Ursa Major is larger than can be
accounted for by its depth. After taking into account uncertainties
in the observations and a cluster depth of 0.17 mag, an intrinsic
scatter in the magnitude of an individual galaxy of order 0.36 mag is
needed.

\item The scatter in the Ursa Major cluster is slightly smaller than
in Virgo.  This is evidence for subclustering in Virgo, assuming that
both clusters are spherical.

\item The slope of the \tf\ in total H-magnitudes is $10.2\pm 0.6$
(for $H^c_{-0.5}$ magnitudes and $\Delta V _{20}^c$ velocity widths).
Since the stellar $M/L$ in this band is rather insensitive to
metallicity, this slope corresponds to
$M \propto V^{4.1 \pm 0.3}$, in
agreement with AHM but not with PT.

\item Dust absorption might cause scatter of $\sim$0.1~mag, but we
see no direct evidence for it.

\item The \tf\  in Ursa Major appears to turn over at large velocity widths.

\item  Distance estimates may be slightly improved if the central
surface bright\-ness of
the galaxy disk is ad\-ded as a third parameter.  This means that the
\tf\ can better be replaced by a magnitude -- velocity width --
surface brightness plane.  The usefulness of this relation should
be tested on an independent sample.

\item There is evidence for a background sub-cluster superposed on
the northern part of the Ursa Major cluster at a heliocentric
velocity $\ga$1000\kms.  If this sub-cluster is real, the intrinsic
scatter in the \tf\ might be quite small.

\end{itemize}

\section*{ACKNOWLEDGEMENTS}

We thank J.~Huchra for permission to use the redshift survey data and
C.~Clemens and E.~Falco for facilitating our access.  We thank
M.~Franx for providing the GALPHOT program, for stimulating
discussions about spiral galaxies, and for valuable comments on the
manuscript. We are indebted to S.~Raychaudhury for taking some of the
images for this project and for showing us his Coma data prior to
publication.  J.~Geary, W.~Wyatt, and C.~Hughes constructed major
parts of SONIC. We thank D.~Mink for help with the software and
B.~van't Sant and T.~Groner for help at the telescope. Part of this
work was done during a visit of R.F.P.\  at the CfA supported by the
Smithsonian Short Term Visitor Program.  This research was partially
supported by grants from the Scholarly Studies Fund of the
Smithsonian Institution and was greatly facilitated by use of the
NASA/IPAC Extragalactic Database (NED) which is operated by the Jet
Propulsion Laboratory, California Institute of Technology, under
contract with the National Aeronautics and Space Administration.




\onecolumn
\begin{texttable}
{BASIC DATA FOR SELECTED URSA MAJOR GALAXIES}
\label{tab:sam}
\begin{tabular}{lrcccccc}
\hline\hline
\setcounter{colcounter}{0}
\0{Galaxy\rule{0pt}{15pt} }
&   \multicolumn{1}{c}{$V_h$}
&   $d_c$
&   $\log(a/b)$
&   Type
&   $\Delta V_{20}$
&   $\pm$
&   $\log\Delta V_{20}^c$ \\
\multicolumn{1}{c}{\col} &
\multicolumn{1}{c}{\col} &
\multicolumn{1}{c}{\col} &
\multicolumn{1}{c}{\col} &
\multicolumn{1}{c}{\col} &
\multicolumn{1}{c}{\col} &
\multicolumn{1}{c}{\col} &
\multicolumn{1}{c}{\col} \\
\hline
NGC~3718 &  987  &  5.74 &  0.31 & 1 & 470 & 3 & 2.726 \\
NGC~3726 &  861  &  4.27 &  0.16 & 5 & 284 & 4 & 2.589 \\
NGC~3729 & 1096  &  5.66 &  0.17 & 1 & 214 &21 & 2.459 \\
NGC~3769 &  724  &  3.23 &  0.50 & 3 & 272 & 9 & 2.452 \\
NGC~3877 &  903  &  2.09 &  0.63 & 5 & 368 & 7 & 2.572 \\
NGC~3893 &  977  &  1.32 &  0.21 & 5 & 302 & 4 & 2.578 \\
NGC~3917 &  975  &  3.34 &  0.61 & 6 & 293 & 5 & 2.475 \\
NGC~3949 &  786  &  0.89 &  0.24 & 4 & 276 & 7 & 2.522 \\
NGC~3953 & 1037  &  3.76 &  0.30 & 4 & 425 & 7 & 2.687 \\
NGC~3972$^a$ &  831  &  6.72 &  0.56 & 4 & 263 & 8 & 2.433 \\
NGC~3985 &  946  &  0.27 &  0.19 & 9 & 165 &11 & 2.327 \\
NGC~3992$^a$ & 1051  &  4.77 &  0.21 & 4 & 475 & 3 & 2.774 \\
NGC~4010 &  905  &  1.39 &  0.73 & 7 & 269 & 7 & 2.432 \\
NGC~4013 &  835  &  4.67 &  0.71 & 3 & 407 & 8 & 2.613 \\
NGC~4085 &  750  &  2.24 &  0.55 & 5 & 292 & 7 & 2.478 \\
NGC~4088 &  752  &  2.41 &  0.41 & 4 & 363 & 6 & 2.590 \\
NGC~4096 &  559  &  1.96 &  0.57 & 5 & 325 & 4 & 2.523 \\
NGC~4100 & 1080  &  1.83 &  0.48 & 4 & 411 & 7 & 2.633 \\
NGC~4102 &  862  &  4.36 &  0.24 & 3 & 315 & 8 & 2.580 \\
NGC~4142 & 1141  &  4.91 &  0.27 & 7 & 187 & 8 & 2.339 \\
NGC~4157$^a$ &  771  &  2.98 &  0.70 & 3 & 413 & 7 & 2.620 \\
NGC~4183$^a$ &  934  &  5.76 &  0.83 & 6 & 249 & 5 & 2.396 \\
NGC~4217 & 1032  &  3.62 &  0.53 & 3 & 431 & 6 & 2.650 \\
NGC~4218$^b$ &  725  &  3.24 &  0.22 & 1 & 160 & 8 & 2.296 \\
NGC~4389 &  717  &  5.84 &  0.29 & 4 & 188 & 5 & 2.337 \\
UGC~6667      &  971  &  3.72 &  0.89 & 6  & 189 & 8  & 2.276 \\
UGC~6917$^c$  & 909   & 1.82  & 0.24  &  9 & 195 &  7 & 2.371 \\
UGC~6923$^a$ & 1065  &  4.56 &  0.38 & 10\0 & 173 & 4 & 2.274 \\
UGC~6983 & 1068  &  4.12 &  0.16 & 6 & 194 & 5 & 2.424 \\
\hline
\end{tabular}

\begin{minipage}[t]{\textwidth}
\noindent
\null$^a$No images obtained; magnitudes from A82.
\null$^b$Not observed.
\null$^c$Bright foreground star; magnitudes could not be derived.

\noindent
{\sc Key to columns---}
(1) Galaxy name;
(2) Heliocentric velocity from CfA redshift survey (Huchra \etal\ 1983 and
private communication);
(3) Distance from center of UMa cluster;
(4) Log(axis ratio), from RC3;
(5) Galaxy type, from RC3;
(6) \hi\ width at 20\%  of the peak (from RC3);
(7) Uncertainty in (6);
(8) Logarithm of inclination-corrected velocity width.
\end{minipage}
\end{texttable}

\textheight=9.2in
\clearpage
\begin{texttable}{REJECTED URSA MAJOR GALAXIES}
\label{tab:exc}
\begin{tabular}{lr}
\hline\hline
\begin{tabular}[t]{crccc}
Galaxy
&   \multicolumn{1}{c}{$V_h$}
&   $d_c$
&   $\log(a/b)$
&   Type
\\
 \multicolumn{1}{c}{\col}
&\multicolumn{1}{c}{\col}
&\multicolumn{1}{c}{\col}
&\multicolumn{1}{c}{\col}
&\multicolumn{1}{c}{\col}
\\
\hline
\multicolumn{5}{l}{Galaxies with type $\le$ 0 or peculiar:} \\
NGC~3870 &  750  & 2.34 & 0.09 & $-$2\0 \\
NGC~3896 &  906  & 1.27 & 0.15 &    0   \\
NGC~3931 &  928  & 3.50 & 0.09 & $-$3\0 \\
NGC~3990 &  705  & 6.85 & 0.24 &    0   \\
NGC~3998 & 1028  & 6.85 & 0.08 &    0   \\
NGC~4026 &  944  & 2.40 & 0.61 & $-$2\0 \\
NGC~4111 &  806  & 5.85 & 0.67 & $-$1\0 \\
NGC~4117 &  958  & 5.85 & 0.31 & $-$2\0 \\
NGC~4138 &  835  & 5.45 & 0.18 & $-$1\0 \\
NGC~4143 &  966  & 6.52 & 0.20 & $-$2\0 \\
NGC~4220 &  954  & 3.36 & 0.45 & $-$1\0 \\
NGC~4346 &  762  & 4.86 & 0.42 & $-$2\0 \\
NGC~4460 &  558  & 6.82 & 0.53 & $-$1\0 \\
UGC~6805 & 1033  & 6.66 & 0.12 & \0E?   \\
UGC~6818 &  803  & 2.98 & 0.33 &   SB?  \\
\end{tabular} &
\begin{tabular}[t]{crccc}
\setcounter{colcounter}{0}
Galaxy
&   \multicolumn{1}{c}{$V_h$}
&   $d_c$
&   $\log(a/b)$
&   Type
\\
\multicolumn{1}{c}{\col}
&\multicolumn{1}{c}{\col}
&\multicolumn{1}{c}{\col}
&\multicolumn{1}{c}{\col}
&\multicolumn{1}{c}{\col}
\\
\hline
\multicolumn{5}{l}{Galaxies with inclination $<$ 45$^{\circ}$:} \\
NGC~3906 &  962 & 1.16 & 0.05 &  7     \\
NGC~3913 &  953 & 6.80 & 0.01 &  7     \\
NGC~3928 &  974 & 0.79 & 0.00 &  3     \\
NGC~3924 & 1003 & 1.55 & 0.04 &  9     \\
NGC~3938 &  812 & 4.53 & 0.04 &  5     \\
NGC~4051 &  710 & 4.24 & 0.13 &  4     \\
UGC~6628 &  849 & 3.93 & 0.00 &  9     \\
UGC~6713 &  896 & 2.03 & 0.14 &  9     \\
UGC~6922 &  892 & 2.21 & 0.09 &\0S?    \\
UGC~6956 &  916 & 2.33 & 0.03 &  9     \\
UGC~6962  & 809 & 5.88 & 0.09 &  6     \\
\end{tabular}
\\
\multicolumn{2}{c}{
\begin{tabular}[t]{crcccccc}
\hline
\setcounter{colcounter}{0}
Galaxy
&   \multicolumn{1}{c}{$V_h$}
&   $d_c$
&   $\log(a/b)$
&   Type
&   $\Delta V_{20}$
&   $\pm$
&   $\log\Delta V_{20}^c$ \\
\multicolumn{1}{c}{(1)} &
\multicolumn{1}{c}{(2)} &
\multicolumn{1}{c}{(3)} &
\multicolumn{1}{c}{(4)} &
\multicolumn{1}{c}{(5)} &
\multicolumn{1}{c}{(6)} &
\multicolumn{1}{c}{(7)} &
\multicolumn{1}{c}{(8)} \\
\hline
\multicolumn{8}{l}{Galaxies without \hi\ data, without \hi,
or with $W_{20}^c < 187.5$\kms\rule{0pt}{14pt}} \\
\0NGC~3769A & 791 & 3.23 & 0.38 &  9 &     &    &       \\
NGC~3782 &  740  &  3.63 &  0.19 & 6 & 132 & 7 & 2.230 \\
UGC~6399  & 779 & 5.72 & 0.55 &  9 & 166 & 16 & 2.233 \\
UGC~6446  & 646 & 6.78 & 0.19 &  7 & 143 &  6 & 2.272 \\
UGC~6773  & 925 & 1.83 & 0.29 & 10\0 &     &    &       \\
UGC~6840  &1016 & 3.57 & 0.50 &  9 & 149 &  6 & 2.190 \\
UGC~6894 &  767  &  6.05 &  0.78 & 6 & 155 & 9 & 2.191 \\
UGC~6930  & 776 & 0.68 & 0.20 &  7 & 133 &  8 & 2.228 \\
UGC~6969 & 1113 & 4.83 & 0.50 & 10\0 & 157 & 10 & 2.213 \\
UGC~6973  & 704 & 5.90 & 0.35 &  2 &     &    &       \\
UGC~7089  & 778 & 5.73 & 0.69 &  8 & 153 &  9 & 2.190 \\
UGC~7176  & 859 & 2.84 & 0.49 & 10\0 & 111 &  9 & 2.064 \\
UGC~7218  & 791 & 4.43 & 0.29 & 10\0 & 107 &  8 & 2.090 \\
UGC~7301  & 712 & 4.31 & 0.88 &  7 & 144 &  9 & 2.158 \\
Mark~1460  & 768 & 1.02 &      &    &     &    &       \\
\0\0115400+4836 & 886 & 0.28 &      &    &     &    &       \\
\0\0115640+5059 & 963 & 2.14 &      &    &     &    &       \\
\end{tabular}
}
\\\hline
\end{tabular}
\begin{minipage}[t]{454pt}
\smallskip
\noindent\small
{\sc Key to columns---}
(1) Galaxy name;
(2) Heliocentric velocity from CfA redshift survey (Huchra \etal\ 1983 and
private communication);
(3) Distance from center of UMa cluster;
(4) Logarithm of axis ratio from RC3;
(5) Galaxy type from RC3;
(6) \hi\ width at 20\%  of the peak (from RC3);
(7) Uncertainty in (6);
(8) Logarithm of inclination-corrected velocity width.
\end{minipage}
\end{texttable}

\pagebreak
\begin{texttable}
{PHOTOMETRIC RESULTS}
\label{tab:mag}
\begin{tabular}{lrrrrrrrrr}
\hline\hline
\multicolumn{1}{c}{Galaxy\rule{0pt}{15pt}}
&  \multicolumn{1}{c}{$D_0$ }
&  \multicolumn{1}{c}{$H^c_{-0.5}$}
&  \multicolumn{1}{c}{$\pm$}
&  \multicolumn{1}{c}{$H^e_{-0.5}$}
& \multicolumn{1}{c}{$\pm$}
& \multicolumn{1}{c}{$H_{19}$}
& \multicolumn{1}{c}{$\pm$}
& \multicolumn{1}{c}{$H_{20}^{0.7}$}
& \multicolumn{1}{c}{$\pm$ }\\
\multicolumn{1}{c}{\col} &
\multicolumn{1}{c}{\col} &
\multicolumn{1}{c}{\col} &
\multicolumn{1}{c}{\col} &
\multicolumn{1}{c}{\col} &
\multicolumn{1}{c}{\col} &
\multicolumn{1}{c}{\col} &
\multicolumn{1}{c}{\col} &
\multicolumn{1}{c}{\col} &
\multicolumn{1}{c}{\col} \\
\hline
NGC~3718 & 476.6 &  8.29 & 0.19 & 8.43 &0.13 & 8.62 &0.08&  8.57& 0.09 \\
NGC~3726 & 353.3 &  8.85 & 0.30 & 8.98 &0.22 & 9.32 &0.16&  9.21& 0.18 \\
NGC~3729 & 177.1 &  9.53 & 0.10 & 9.77 &0.05 & 9.41 &0.06&  9.40& 0.06 \\
NGC~3769 & 157.8 &  9.94 & 0.09 &10.42 &0.05 &10.10 &0.06& 10.16& 0.06 \\
NGC~3782 &  97.3 & 11.69 & 0.05 &      &     &      &    &      &      \\
NGC~3877 & 250.1 &  8.67 & 0.06 & 9.28 &0.05 & 8.66 &0.05&  8.82& 0.05 \\
NGC~3893 & 244.4 &  8.72 & 0.08 & 8.92 &0.06 & 8.79 &0.07&  8.79& 0.07 \\
NGC~3917 & 233.4 &  9.93 & 0.09 &10.72 &0.05 &10.61 &0.05& 10.25& 0.05 \\
NGC~3949 & 165.3 &  9.37 & 0.05 & 9.63 &0.05 & 9.29 &0.05&  9.30& 0.05 \\
NGC~3953 & 361.5 &  7.91 & 0.09 & 8.17 &0.06 & 7.99 &0.07&  8.02& 0.07 \\
NGC~3972 & 194.2 & 10.46 & 0.05 &      &     &      &    &      &      \\
NGC~3985 &  70.5 & 11.35 & 0.05 &11.56 &0.05 &11.15 &0.05& 10.94& 0.05 \\
NGC~3992 & 424.8 &  7.95 & 0.05 &      &     &      &    &      &      \\
NGC~4010 & 185.4 & 10.44 & 0.10 &11.05 &0.05 &10.97 &0.05& 10.71& 0.05 \\
NGC~4013 & 233.4 &  8.53 & 0.11 & 8.90 &0.05 & 8.26 &0.08&  8.20& 0.08 \\
NGC~4085 & 137.5 &  9.99 & 0.05 &10.47 &0.05 &10.07 &0.05& 10.15& 0.05 \\
NGC~4088 & 293.9 &  8.33 & 0.13 & 8.81 &0.08 & 8.27 &0.12&  8.41& 0.10 \\
NGC~4096 & 307.7 &  8.84 & 0.08 & 9.36 &0.05 & 9.10 &0.05&  9.08& 0.05 \\
NGC~4100 & 261.9 &  8.83 & 0.10 & 9.36 &0.06 & 8.88 &0.07&  9.06& 0.07 \\
NGC~4102 & 177.1 &  8.42 & 0.05 & 8.56 &0.05 & 8.31 &0.05&  8.39& 0.05 \\
NGC~4142 & 128.3 & 12.10 & 0.07 &12.47 &0.05 &16.07 &0.08& 13.16& 0.05 \\
NGC~4157 & 307.7 &  8.32 & 0.05 &      &     &      &    &      &      \\
NGC~4183 & 212.9 & 10.64 & 0.05 &      &     &      &    &      &      \\
NGC~4217 & 268.0 &  8.59 & 0.05 & 8.99 &0.05 & 8.65 &0.05&  8.65& 0.05 \\
NGC~4389 & 150.7 & 10.45 & 0.05 &10.69 &0.05 &10.89 &0.05& 10.51& 0.05 \\
UGC~6667 & 143.9 & 12.14 & 0.07 &13.13 &0.05 &14.11 &0.05& 13.22& 0.05 \\
UGC~6923 & 104.3 & 12.04 & 0.05 &      &     &      &    &      &      \\
UGC~6983 & 222.9 & 11.55 & 0.20 &11.77 &0.05 &14.99 &0.06& 13.45& 0.07 \\
\multicolumn{8}{l}{\rule{0pt}{15pt}Extra galaxy, not in sample:}\\
UGC~6894 &  62.8 & 13.71 & 0.08 &14.74 &0.06 &$<$16.00 &0.10& 14.76& 0.06 \\
\hline
\end{tabular}

\begin{minipage}[t]{\textwidth}
\noindent
{\sc Key to columns---}
(1) Galaxy name;
(2) Isophotal diameter $D_0$, corrected for Galactic extinction and
inclination;
(3) circular magnitude inside $0.316~D_0$ (derived from A82
if other magnitudes are missing);
(4) uncertainty in (3);
(5) elliptical magnitude inside $0.316~D_0$;
(6) uncertainty in (5);
(7) elliptical magnitude inside $D_{19,i}$;
(8) uncertainty in (7);
(9) elliptical magnitude inside $0.7 D_{20,i}$;
(10) uncertainty in (9).
\end{minipage}
\end{texttable}

\pagebreak
\begin{texttable}
{BULGE-DISK DECOMPOSITION}
\label{tab:buldis}
\tablehead{
\hline\hline
\multicolumn{1}{c}{Galaxy}
& $\epsilon_B$
& $\epsilon_D$
& $\phi_D$
& $H_T$
& $B/D$
& $h_D$
& $H(0)$
& $H_D$ \\
\multicolumn{1}{c}{\col} &
\multicolumn{1}{c}{\col} &
\multicolumn{1}{c}{\col} &
\multicolumn{1}{c}{\col} &
\multicolumn{1}{c}{\col} &
\multicolumn{1}{c}{\col} &
\multicolumn{1}{c}{\col} &
\multicolumn{1}{c}{\col} &
\multicolumn{1}{c}{\col} \\
\hline}
\tabletail{\hline}
\begin{supertabular}{lrrrrrrrr}
NGC~3718 & 0.07 & 0.17 & 112 & 8.05 & 0.41 & 26.8 & 17.41 & 8.43 \\
NGC~3726 & 0.30 & 0.47 & 112 & 8.22 & 0.05 & 37.6 & 17.50 & 8.27 \\
NGC~3729 & 0.18 & 0.36 &  84 & 8.64 & 0.03 & 21.9 & 16.93 & 8.67 \\
NGC~3769 & 0.20 & 0.65 &  63 & 9.45 & 0.10 & 17.5 & 16.66 & 9.55 \\
NGC~3877 & 0.21 & 0.77 & 123 & 8.26 & 0.06 & 28.9 & 16.06 & 8.32 \\
NGC~3893 & 0.15 & 0.38 & 42  & 8.24 & 0.46 & 26.5 & 17.29 & 8.65 \\
NGC~3917 & 0.30 & 0.73 & 167 & 8.67 & 0.02 & 52.6 & 17.92 & 8.69 \\
NGC~3949 & 0.37 & 0.44 &  32 & 8.73 & 0.26 & 20.1 & 16.90 & 8.98 \\
NGC~3953 & 0.23 & 0.47 & 102 & 7.43 & 0.11 & 34.6 & 16.59 & 7.54 \\
NGC~3985 & 0.33 & 0.34 &  10 &10.33 & 0.11 & 11.1 & 17.25 &10.44 \\
NGC~4010 & 0.50 & 0.84 & 155 & 9.50 & 0.03 & 35.7 & 17.35 & 9.54 \\
NGC~4013 & 0.15 & 0.84 & 154 & 7.74 & 0.09 & 34.1 & 15.54 & 7.83 \\
NGC~4085 & 0.15 & 0.67 & 165 & 9.11 & 0.06 & 20.5 & 16.56 & 9.17 \\
NGC~4088 & 0.30 & 0.61 & 143 & 7.58 & 0.03 & 39.3 & 16.61 & 7.62 \\
NGC~4096 & 0.25 & 0.72 & 108 & 7.99 & 0.08 & 50.5 & 17.25 & 8.07 \\
NGC~4100 & 0.25 & 0.67 &  75 & 8.15 & 0.05 & 36.1 & 16.82 & 8.20 \\
NGC~4102 & 0.11 & 0.27 & 131 & 7.96 & 0.56 & 18.9 & 16.52 & 8.44 \\
NGC~4142 &      & 0.33 &  89 &11.23 & 0.00 & 16.0 & 18.85 &11.23 \\
NGC~4217 & 0.42 & 0.85 & 139 & 7.84 & 0.03 & 47.7 & 16.24 & 7.87 \\
NGC~4389 & 0.70 & 0.79 &  14 & 9.27 & 0.07 & 39.4 & 17.67 & 9.35 \\
UGC~6667 & 0.00 & 0.84 & 177 &11.00 & 0.03 & 32.7 & 18.65 &11.03 \\
UGC~6983 &      & 0.30 & 180 &11.00 & 0.01 & 17.0 & 18.43 &11.01 \\
\multicolumn{8}{l}{\rule{0pt}{15pt}Extra galaxy, not in sample:}\\
UGC~6894 &      & 0.86 &   2 &12.21 & 0.00 & 23.0 & 18.91 &12.21 \\
\hline
\end{supertabular}
\noindent

\begin{minipage}[t]{\textwidth}
{\sc Key to columns---}
(1) Galaxy name;
(2) Ellipticity of bulge;
(3) Ellipticity of disk;
(4) Position angle  of galaxy disk (north through east);
(5) Total $H$ magnitude;
(6) Bulge to disk ratio (in luminosity);
(7) $H$ scale length of disk;
(8) Central disk surface brightness;
(9) Total magnitude of disk.
\end{minipage}
\end{texttable}

\pagebreak
\begin{texttable}
{LEAST-SQUARE FIT RESULTS --- INGREDIENTS OF TFR}
\label{tab:lsq}
\begin{tabular}{lcccccccc}
\hline\hline
\multicolumn{1}{c}{Magnitude}\rule{0pt}{15pt}
& \multicolumn{1}{c}{Non-rot.}
& \multicolumn{1}{c}{Incl.}
& \multicolumn{1}{c}{Sample}
& \multicolumn{1}{c}{$N_{gal}$}
& $\chi^2_{red}$
& $\Delta$(mag)
& Slope
& Intcpt.\\
\multicolumn{1}{c}{Type}
& \multicolumn{1}{c}{Correction}
&  &  &  &  &  &  &  \\
\multicolumn{1}{c}{\col} &
\multicolumn{1}{c}{\col} &
\multicolumn{1}{c}{\col} &
\multicolumn{1}{c}{\col} &
\multicolumn{1}{c}{\col} &
\multicolumn{1}{c}{\col} &
\multicolumn{1}{c}{\col} &
\multicolumn{1}{c}{\col} &
\multicolumn{1}{c}{\col} \\
\hline
\multicolumn{9}{l}{Changing magnitude type:} \\
{}~~~$H^c_{-0.5}$   & no &  RC3  & SONIC  &  22 & 4.61  & 0.41 & 0.096 & 2.564
\\
{}~~~$H^e_{-0.5}$   & no & RC3   & SONIC  &  22 & 5.46  & 0.49 & 0.085 & 2.603
\\
{}~~~$H_{19}$       & no & RC3   & SONIC  &  22 &11.31  & 1.35 & 0.055 & 2.568
\\
{}~~~$H_{20}^{0.7}$ & no & RC3   & SONIC  &  22 & 8.35  & 0.84 & 0.073 & 2.571
\\
{}~~~$H_D$          & no & RC3   & SONIC  &  22 & 7.55  & 0.54 & 0.110 & 2.493
\\
{}~~~$H_T$          & no & RC3   & SONIC  &  22 & 6.87  & 0.52 & 0.108 & 2.484
\\
\multicolumn{9}{l}{Changing type of velocity width:} \\
{}~~~$H_T$          &yes & RC3   & SONIC  &  22 & 7.29  & 0.51 & 0.125 & 2.422
\\
{}~~~$H^c_{-0.5}$   & no &  RC3  & Table~1&  27 & 4.32  & 0.40 & 0.098 & 2.564
\\
{}~~~$H^c_{-0.5}$   &yes & RC3   & Table~1&  27 & 4.64  & 0.39 & 0.112 & 2.515
\\
\multicolumn{9}{l}{Changing inclinations:} \\
{}~~~$H^c_{-0.5}$   & no & IR    & SONIC  &  22 & 9.29  & 0.64 & 0.110 & 2.563
\\
{}~~~$H_T$          & no & IR    & SONIC  &  22 &12.92  & 0.70 & 0.122 & 2.470
\\
\hline
\end{tabular}

\begin{minipage}[t]{\textwidth}
\noindent
{\sc Key to columns---}
(1) Type of H-magnitude (Tables~\ref{tab:mag}, \ref{tab:buldis});
(2) Correction for non-rotational motion ($W_R$);
(3) Source of inclinations;
(4) Sample (The ``SONIC'' sample
consists of the galaxies in Table~\ref{tab:sam}
excluding the five that lack images.);
(5) Number of galaxies in sample;
(6) Reduced $\chi^2$ of fit;
(7) Additional uncertainty in the $H$~magnitudes necessary to
bring $\chi^2$ down to unity;
(8) and (9) best slope $a$ and intercept $b$ for $\log \Delta V =
-a(H - 9.0) + b$.
\end{minipage}
\end{texttable}

\begin{texttable}
{MULTIPLE PARAMETER FITS}
\label{tab:bivar}
\begin{tabular}{lccccl}
\hline\hline
\multicolumn{1}{c}{Mag.}
& \multicolumn{1}{c}{Sample}
& \multicolumn{1}{c}{$N_{gal}$}
& \multicolumn{1}{c}{$\Delta$(mag)}
& \multicolumn{1}{c}{$\Delta$(mag)}
& \multicolumn{1}{c}{Relation derived} \\
\multicolumn{1}{c}{Type}
&
&
&\multicolumn{1}{c}{2 par}
&\multicolumn{1}{c}{3 par}
&\multicolumn{1}{c}{($H =$)} \\
\multicolumn{1}{c}{\col} &
\multicolumn{1}{c}{\col} &
\multicolumn{1}{c}{\col} &
\multicolumn{1}{c}{\col} &
\multicolumn{1}{c}{\col} &
\multicolumn{1}{c}{\col} \\
\hline
\rule{0pt}{15pt}~~~$H^c_{-0.5}$   &   SONIC &   22 & 0.41 &  0.35 &
$21.41 - 7.74\log\Delta V^c + 0.44 H(0)$ \\
{}~~~$H^c_{-0.5}$   & Omit NGC~4389 &   21 & 0.37 &  0.32 &
$23.16 - 8.29\log\Delta V^c + 0.43 H(0)$ \\
{}~~~$H_T$   &   SONIC &   22 & 0.52 &  0.45 &
$15.96 - 6.10\log\Delta V^c + 0.48 H(0)$ \\
{}~~~$H_T$   &   Omit NGC~4389 &   21 & 0.47 &  0.41 &
$18.11 - 6.77\log\Delta V^c + 0.45 H(0)$ \\
\hline
\end{tabular}

\begin{minipage}[t]{\textwidth}
\noindent
{\sc Key to columns---}
(1) Type of $H$~magnitude;
(2) Sample;
(3) Number of galaxies in sample;
(4) Additional uncertainty in the $H$~magnitudes necessary to
bring $\chi^2$ down to unity with uncertainties only in
magnitudes and velocity widths;
(5) Same as (4) but with uncertainties in  $H(0)$ as well;
(6) Best fitting plane.
\end{minipage}
\end{texttable}

\pagebreak
\begin{texttable}
{LEAST-SQUARE FIT RESULTS --- CHANGING SAMPLE}
\label{tab:pt}
\begin{tabular}{ccccccccc}
\hline\hline
\multicolumn{1}{c}{Velocity}\rule{0pt}{15pt}
& \multicolumn{1}{c}{Non-rot.}
& \multicolumn{1}{c}{Incl.}
& \multicolumn{1}{c}{Sample}
& \multicolumn{1}{c}{$N_{gal}$}
& $\chi^2_{red}$
& $\Delta$(mag)
& Slope
& Intcpt.\\
\multicolumn{1}{c}{Widths}
& \multicolumn{1}{c}{Correction}
&  &  &  &  &  &  &  \\
\multicolumn{1}{c}{\col} &
\multicolumn{1}{c}{\col} &
\multicolumn{1}{c}{\col} &
\multicolumn{1}{c}{\col} &
\multicolumn{1}{c}{\col} &
\multicolumn{1}{c}{\col} &
\multicolumn{1}{c}{\col} &
\multicolumn{1}{c}{\col} &
\multicolumn{1}{c}{\col} \\
\hline
\multicolumn{7}{l}{Baseline:} \\
RC3        & no &  RC3  & Table~1&  27 & 4.32  & 0.40 & 0.098 & 2.564 \\
\multicolumn{9}{l}{Comparison with PT:} \\
PT         & yes&  PT   & PT     &  18 & 2.84  & 0.20 & 0.106 & 2.516 \\
PT         & yes&  PT   & PT$+$2 &  20 & 2.64  & 0.20 & 0.106 & 2.516 \\
RC3        & yes&  PT   & PT$+$2 &  20 & 2.78  & 0.20 & 0.109 & 2.513 \\
RC3        & no &  PT   & PT$+$2 &  20 & 2.54  & 0.19 & 0.095 & 2.563 \\
RC3        & no &  RC3  & PT$+$2 &  20 & 2.59  & 0.26 & 0.097 & 2.566 \\
RC3        & no &  RC3  & omit 3782&  19 & 2.50  & 0.24 & 0.096 & 2.566 \\
\multicolumn{9}{l}{Omit NGC~3718:} \\
RC3        & no &  RC3  & Table~1&  26 & 3.99  & 0.38 & 0.096 & 2.562 \\
\multicolumn{9}{l}{Omit NGC~4389:} \\
RC3        & no &  RC3  &Table~1 &  26 & 4.00  & 0.37 & 0.096 & 2.565 \\
\multicolumn{9}{l}{Omit NGC~3718 and NGC~4389:} \\
RC3        & no &  RC3  & Table~1&  25 & 3.65  & 0.34 & 0.094 & 2.563 \\
\multicolumn{9}{l}{Omit NGC~4096:} \\
RC3        & no &  RC3  &Table~1 &  26 & 3.58  & 0.38 & 0.099 & 2.570 \\
\multicolumn{7}{l}{Bright galaxies: $\log\Delta V^c ~>~ 2.43$ :} \\
RC3        & no &  RC3  & Table~1&  20 & 4.44  & 0.31 & 0.104 & 2.565 \\
\multicolumn{9}{l}{Galaxies with $600 < V_H < 1025$\kms:} \\
RC3        & no &  RC3  & Table~1&  18 & 2.20  & 0.26 & 0.095 & 2.557 \\
\multicolumn{9}{l}{Galaxies with $600 < V_H <  985$\kms:} \\
RC3        & no &  RC3  & Table~1&  17 & 1.32  & 0.12 & 0.092 & 2.553 \\
\hline
\end{tabular}

\begin{minipage}[t]{\textwidth}
\noindent
{\sc Key to columns---}
(1) Source of velocity widths;
(2) Correction for non-rotational motion ($W_R$);
(3) Source of inclinations;
(4) Sample;
(5) Number of galaxies in sample;
(6) Reduced $\chi^2$ of fit;
(7) Additional uncertainty in the $H$~magnitudes necessary to
bring $\chi^2$ down to unity;
(8) and (9) best slope $a$ and intercept $b$ for $\log \Delta V =
-a(H - 9.0) + b$. For the baseline, the uncertainty in the slope
is 0.006, and the uncertainty   in the intercept is 0.008.
\end{minipage}
\end{texttable}

\pagebreak
\begin{center}
\bf FIGURE CAPTIONS
\end{center}
\nopagebreak
\noindent
\begin{enumerate}
\renewcommand{\makelabel}{\refstepcounter{figure}Fig.~}
\renewcommand{\labelenumi}{\arabic{enumi}---}
\labelsep=0pt
\item\label{fig:typ}
Images of three galaxies in the sample, NGC 3877, one of our largest
galaxies, NGC~3729, an intermediate size galaxy, and UGC~6667, one of
our smallest galaxies. The three are shown on the same
angular and intensity scales, so the lower central surface brightness
of the fainter galaxies can be seen.

\item\label{fig:a82}
Comparison between SONIC magnitudes and magnitudes from single-detector
aper\-ture photo\-metry (A82). For this figure only, SONIC magnitudes
are measured in circular apertures of diameter $A$ such that $\log(A/D_1)
= -0.5$.

\item\label{fig:ell}
Tully-Fisher diagram with $a)$ circular, $b)$ elliptical,
$c)$ isophotal, and $d)$ total
magnitudes for the SONIC sample only.  Error bars are shown for the
velocity widths only.  Except for a few galaxies
(Table~\ref{tab:mag}), the error bars on the magnitudes are about
the size of the symbols.  The solid line shows the \tf\ fit to the
baseline sample.  Since the elliptical magnitudes are fainter, they
tend to fall below this line.  The dashed line show the best fits for
the elliptical, isophotal, and total magnitudes.

\item\label{fig:incl}
Comparison between inclinations derived from RC3 axis ratios
(blue photographic measurements) and from infrared images.
For many galaxies the agreement
is very good, but for others the difference can exceed
10\deg.

\item\label{fig:biv}
Tully-Fisher diagram with a third parameter.  Adjusting the magnitude
by the   central surface brightness of the disk gives a flatter slope
and less dispersion in the magnitude axis.  The line shows the best
fit relation from Table~\ref{tab:bivar}.

\item\label{fig:baseline}
Tully-Fisher diagram for the baseline sample.  Galaxies previously
observed by PT are shown with filled symbols while those added are
shown by open symbols.  Vertical error bars are not shown because
they are almost all smaller than the symbols.  The line shows the
best fit relation from Table~\ref{tab:pt}.  NGC numbers are shown for
five of the  galaxies with the largest deviations.

\item\label{fig:ptcom}
Tully-Fisher diagram for the PT sample of galaxies in the Ursa Major
cluster.  Part $a$) uses inclinations from PT, part $b$) from the RC3.

\item\label{fig:vel}
Residuals from the \tf\ as a function of heliocentric radial
velocity.  Galaxies having large residuals are identified with NGC or
UGC numbers.  A positive residual implies that the galaxy is too
faint for its velocity width.

\end{enumerate}

\begin{references}

Aaronson, M., Huchra, J., \& Mould, J. 1979, \apj\vol{229}, 1 (AHM).

Aaronson, M., Mould, J., \& Huchra, J.  1980, \apj\vol{237}, 655 (AMH).

Aaronson, M. \etal\ 1982, \apjs\vol{50}, 241 (A82).

Aaronson, M. \& Mould, J.  1983, \apj\vol{265}, 1.

Aaronson, M., Bothun, G., Mould, J.,
Huchra, J., Schommer, R. A., \& Cornell, M. E., 1986, \apj\vol{302}, 536.

Biviano, A., Giuricin, G., Mandirossian, F., \& Mezzetti, M.,
1990, \apjs\vol{74}, 325.

Bothun, G. D., Mould, J., Schommer, R. A., \& Aaronson, M. 1985,
\apj\vol{291}, 586.

Bothun, G. D. \& Mould, J. R. 1987, \apj\vol{313}, 629.

Bottinelli, L., Fouqu\'e, P., Gouguenheim, L., Paturel, G., \& Teerikorpi, P.
1987, \aap\vol{181}, 1.

Bottinelli, L., Gouguenheim, L., Paturel, G., \& de Vaucouleurs, G.
1983, \aap\vol{118}, 4.

Bottinelli, L., Gouguenheim, L., Paturel, G., \& de Vaucouleurs, G.
1984, \apj\vol{280}, 34.

Bottinelli, L., Gouguenheim, L., Paturel, G., \& Teerikorpi, P.
1988, \apj\vol{328}, 4.

Bottinelli, L., Gouguenheim, L., Fouqu\'e, P., \& Paturel, G. 1990,
\aaps\vol{82}, 391.


de Vaucouleurs, G., de Vaucouleurs, A., \& Corwin, H. G. Jr., 1976,
{\js Second Reference Catalog
of Bright Galaxies}, (Austin:U.~Texas Press) (RC2).

de Vaucouleurs, G., de Vaucouleurs, A., Corwin, H. G., Jr., Buta, R.
J., Paturel, G., \& Fouqu\'{e}, P. 1991, {\js Third Reference Catalog
of Bright Galaxies}, (NY:Springer-Verlag) (RC3).

Elias, J. H., Frogel, J. A., Matthews, K. \&, Neugebauer, G. 1982,
\aj\vol{87}, 1029.

Fouqu\'e, P., Bottinelli, L., Gouguenheim, L., \& Paturel, G., 1990,
\apj\vol{349}, 1.

Franx, M. \& de Zeeuw, P. T.  1992, \apj\vol{392}, L47.

Freedman, W. L., 1990, \apj\vol{355}, L35.

Freeman, K. C. 1970, \apj\vol{160}, 811.

Geller, M. J. \& Huchra, J. P. 1983, \apjs\vol{52}, 61.

Giraud, E., 1986, \apj\vol{301}, 7.

Helou, G., Hoffman, G. L., \& Salpeter, E. E. 1984, \apjs\vol{55}, 433.

Huchra, J., Davis, M., Latham, D., \& Tonry, J. 1983, \apjs\vol{52},
89.

Huchra, J. P. \& Geller, M. J. 1982, \apj\vol{257}, 423.

Huchtmeier, W. K. 1982, \aap\vol{110}, 121.

Jacoby, G. H., Branch, D., Ciardullo, R., Davies, R. L.,
Harris, W. E., Pierce, M. J., Pritchet, C. J., Tonry, J. L., \&
Welch, D. L., 1992, \pasp\vol{104}, 599.

J\o rgensen, I., Franx, M., \& Kj\ae rgaard, P., 1992, \aaps\vol{95}, 489.

Kent, S. M. 1986, \aj\vol{91}, 1301.

Kraan-Korteweg, R. C., Cameron, L. M., \& Tammann, G. A., 1988,
\apj\vol{331}, 620.

Mathewson, D. S., Ford, V. L., \& Buchhorn, M., 1992, \apj\vol{389}, L5.

Peletier, R. F. \&, Willner, S. P. 1991, \apj\vol{382}, 382 (Paper~1).

Peletier, R. F. \&, Willner, S. P. 1992, \aj\vol{103}, 1761.


Pierce, M. J. \& Tully, R. B. 1988, \apj\vol{330}, 579 (PT).

Schommer, R. A., Bothun, G. D., Wil\-liams, T. B., \& Mould, J. R.,
1993, \aj\vol{105}, 97.

Schultz, G. V. \& Wiemer, W. 1975, \aap\vol{43}, 133.

Schwarz, U. J., 1985, \aap\vol{142}, 273.

Teerikorpi, P. 1984, \aap\vol{141}, 407.

\duprule. 1987, \aap\vol{173}, 39.

Tully, R. B. 1987, \apj\vol{321}, 280.

\duprule. 1988, {\js Nearby Galaxies Catalog}, (Cambridge:U.~Press) (NBG).

Tully, R. B. \& Fisher, J. R.,, 1977, \aap\vol{54}, 661.

Tully, R. B. \& Fouqu\'e, P. 1985, \apjs\vol{58}, 67.

Turner, E. L. \& Gott, J. R. 1976, \apjs\vol{32}, 409.

\end{references}
\end{document}